\documentclass[prl,reprint,aps,twocolumn,nopacs,superscriptaddress]{revtex4-2}
\usepackage[utf8]{inputenc}
\usepackage[american,]{babel}
\usepackage[T1]{fontenc}
\usepackage[pdftex]{graphicx}  
\usepackage{graphicx, xcolor}
\usepackage{dcolumn}
\usepackage{bm}
\usepackage{amsmath,amsthm,amssymb}
\usepackage{hyperref}
\usepackage[T1,T2A]{fontenc}
\usepackage{xcolor}
\hypersetup{colorlinks,bookmarksopen,bookmarksnumbered,
    citecolor=blue,
    linkcolor=blue,
    pdfstartview=false,
    urlcolor=blue}
\usepackage{graphicx}
\usepackage{braket}
\usepackage{wrapfig}
\usepackage{soul}
\usepackage{mathtools}
\usepackage{float}
\usepackage{tabularx}
\usepackage{subfigure}
\newcommand{\plink}{P_{\rm link}}

\begin{document}

\title{Anderson localization induced by structural disorder}

\author{Sourav Bhattacharjee}
\affiliation{ICFO – Institut de Ciències Fotòniques, The Barcelona Institute of Science and Technology, Av. Carl Friedrich Gauss 3, 08860 Castelldefels (Barcelona), Spain}
\author{Piotr Sierant}
\affiliation{ICFO – Institut de Ciències Fotòniques, The Barcelona Institute of Science and Technology, Av. Carl Friedrich Gauss 3, 08860 Castelldefels (Barcelona), Spain}
\author{Marek Dudyński}
\affiliation{Qenergy Sp.z oo, Aleja 3 Maja 5/7, 00-401 Warszawa, Poland}
\author{Jan Wehr}
\affiliation{Department of Mathematics, The University of Arizona, Tucson, AZ 85721-0089, United States of America}
\author{Jakub Zakrzewski}
\affiliation{
Instytut Fizyki Teoretycznej, Wydzia\l{} Fizyki, Astronomii i Informatyki Stosowanej,
Uniwersytet Jagiello\'nski, 
L{}ojasiewicza 11, PL-30-348 Krak\'ow, Poland}
\affiliation{Mark Kac Complex Systems Research Center, Uniwersytet Jagiello\'nski, PL-30-348 Krak\'ow, Poland}
\author{Maciej Lewenstein}
\affiliation{ICFO – Institut de Ciències Fotòniques, The Barcelona Institute of Science and Technology, Av. Carl Friedrich Gauss 3, 08860 Castelldefels (Barcelona), Spain}
\affiliation{ICREA, Pg. Lluıs Companys 23, 08010 Barcelona, Spain}

\begin{abstract}
We examine the onset of Anderson localization in three-dimensional systems with structural disorder in form of lattice irregularities and in the absence of any on-site disordered potential. 
Analyzing two models with distinct types of lattice regularities, we show that the Anderson localization transition occurs when the strength of the structural disorder is smoothly increased. Performing finite-size scaling analysis of the results, we show that the transition belongs to the same universality class as regular Anderson localization induced by onsite disorder. 
Our work identifies a new class of structurally disordered lattice models in which destructive interference of matter waves may inhibit transport and lead to a transition between metallic and localized phases.
\end{abstract}

\maketitle

\paragraph{Introduction.} The role of disorder in the localization of quantum systems continues to be a subject of extensive research to date \cite{Kramer93, Evers08, segev13, Nandkishore15, denglevy16, Abanin19, cugliandolo24, cheng24, Sierant24MBLrev}. The phenomenon of Anderson localization (AL) \cite{Anderson58, Abrahams79,  wiersma13}, wherein single-particle wave-functions of non-interacting quantum systems spatially localize in the presence of spatial disorder is fairly well understood~\cite{Evers08} and has been demonstrated experimentally \cite{Billy08, chabe08, lemarie10, lopez12, white20}. In lattice systems, the disorder is traditionally introduced locally by an onsite potential at each lattice site  
sampled randomly from a distribution whose width determines the disorder strength. While any amount of disorder localizes all the eigenstates in one-dimensional (1D) systems, 
disorder exceeding a critical strength is necessary to induce the 
AL transition in three-dimensional (3D) systems (see e.g. \cite{Pasek17} and references therein). Two-dimensional (2D) systems present a marginal case where wavefunctions localize over a spatial extent 
increasing rapidly with the decrease of the disorder strength, rendering AL in 2D difficult to observe experimentally~\cite{Mueller10, Suntajs23}

\begin{figure}
    \centering
    \includegraphics[width=\linewidth]{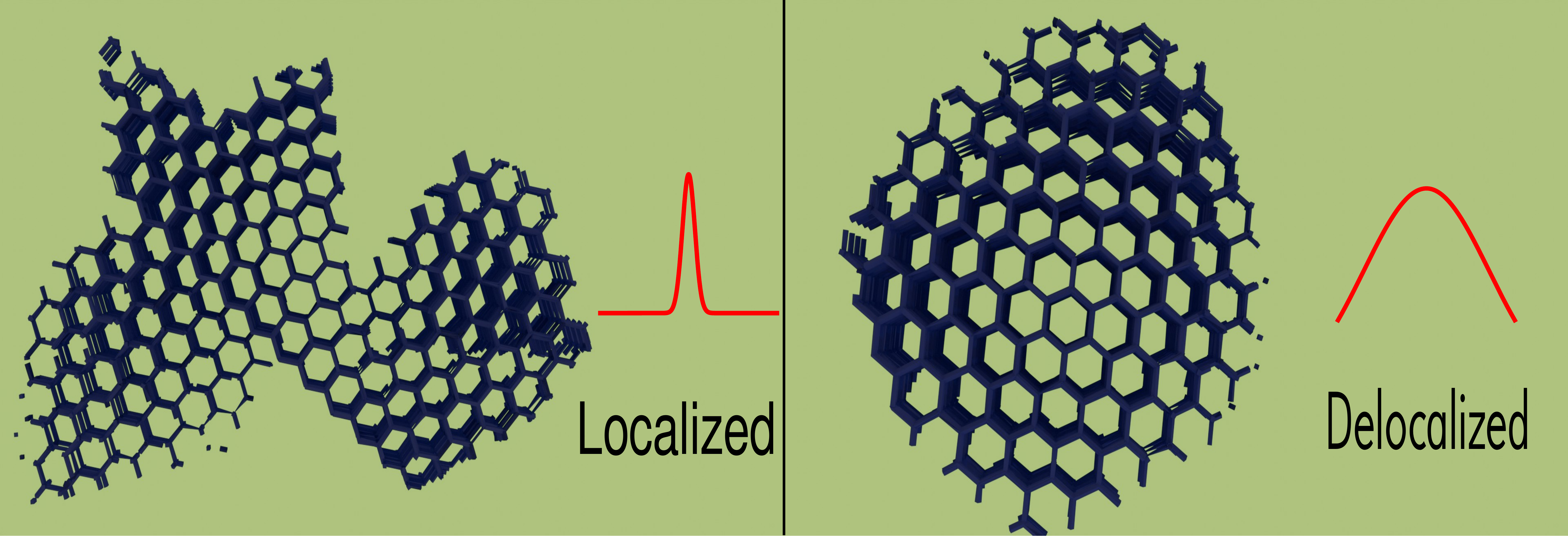}
    \caption{
    Destructive interference arising from structural disorder in the form of lattice irregularities results in AL transition between a localized phase (left panel, strong structural disorder) and a delocalized phase (right panel, weak structural disorder).}
    \label{fig:cartoon_intro}
\end{figure}

Fundamentally, localized eigenstates result from the superposition of propagating wavefunctions with 
randomly distributed phases acquired from scattering off the random onsite potentials~\cite{Mueller10}. One can therefore envisage that the presence of onsite disorder is not a necessary condition for AL; random phases can also be acquired from scattering off irregular surfaces or edges of the system which we shall henceforth refer to as a structural disorder. 
This leads to the following questions:
(\textit{i}) can the structural disorder, analogously to the onsite disorder, lead to an AL transition that separates localized and delocalized phases,  and (\textit{ii}) if  such a localization transition exists, does it belong to the same universality class as the regular AL in the presence of onsite disorder. We note here that AL has previously also been observed with off-diagonal disorder, i.e. disorder introduced directly in the hopping potential of the particle in the lattice \cite{eilmes98, fratini15, martin11,biswas2000}. However, the structural disorder we explore in this work introduces disorder only at the surface of the structures, and hence is long-range correlated throughout the structure.

In this letter, we demonstrate that a 3D AL transition can 
indeed be induced by a smooth increase of strength of the structural disorder. To this end, we examine a tight-binding lattice model with structural disorder introduced by removing lattice sites from a regular 3D honeycomb lattice (2D honeycomb lattices stacked in AA configuration), as shown in Fig.~\ref{fig:cartoon_intro}. The process of removing sites is described by a classical statistical model defined on the chosen lattice.
Focusing on a tight-binding Hamiltonian defined on the largest cluster of the remaining lattice sites, we show that the structural disorder strength can be increased by varying parameters of the underlying statistical model, leading to AL transition. Importantly, the AL transition occurs in the presence of long-range correlated disorder -- as the analysis is restricted to the largest cluster within which there is no disorder in the onsite potential or in the bonds between the sites, the correlations span the full volume of the cluster \cite{suppl}. We investigate the critical properties of the transition and consider an alternative 3D lattice with structural disorder to illustrate the robustness of our findings.

\begin{figure*}
    \centering
    \includegraphics[width=0.4\linewidth]{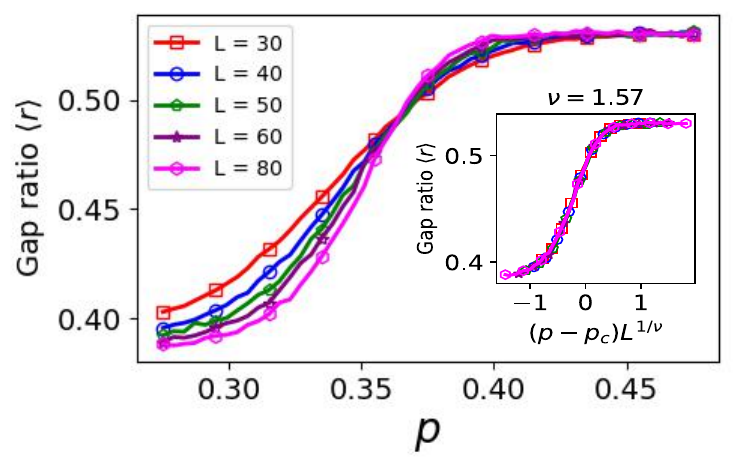}
    \includegraphics[width=0.4\linewidth]{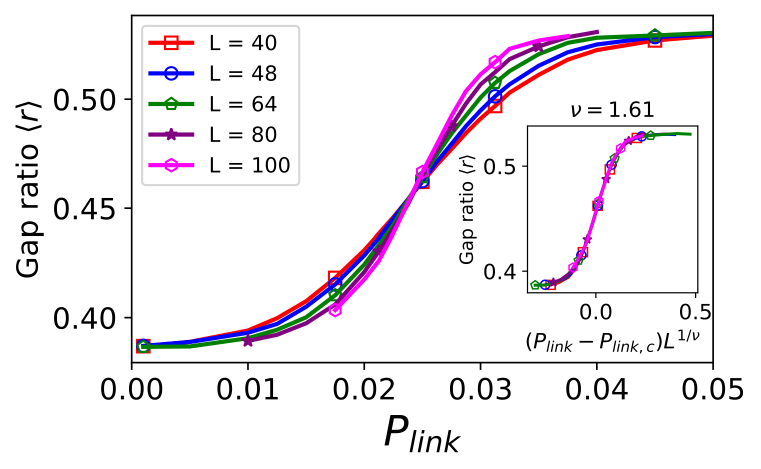}
    \caption{(a) The averaged gap ratio $\langle r\rangle$ captures the AL transition on $\mathcal{C}$ for different system sizes $L$ in the honeycomb model as a function of $p$, where $p$ is the fraction of sites remaining in the lattice. The inset shows a critical scaling collapse with exponent $\nu=1.57$. (b) A similar AL transition is also observed for the link model in which case $P_{link}$ represents the probability of added links in the random spanning tree.}
    \label{fig:gap_ratio}
\end{figure*}

\paragraph{
The honeycomb model.}
We consider a single-particle Hamiltonian
\begin{equation}
 H_{sp}=\sum_{ ( {\bf r},{\bf r'} ) \in \mathcal{T}} c_{\bf r}^\dag c_{\bf r'}
 \label{eq:Ham}
\end{equation}
where $c_{\bf r}$ is an operator annihilating a fermion at site $\bf {r}$ of a lattice $\mathcal{C}$, and $\mathcal{T}$ is a set of the lattice links. When $\mathcal{C}$ is a regular 3D lattice with discrete translational symmetry, and $\mathcal{T}$ comprises links between neighboring lattice sites, the eigenstates of \eqref{eq:Ham} are delocalized Bloch waves~\cite{Bloch29}. Physical processes resulting in removal of lattice sites may lead to breakdown of the translational symmetry and introduce a structural disorder to the system.
A particular experimental scenario which motivates our considerations is Joule heating of biomass~\cite{Bourke07, Tetlow14, Jiang19} which leads to carbonization and graphitization process resulting in creation of an irregular structure of graphene platelets~\cite{Yee23}. To model such irregular structures, we assume that the lattice $\mathcal{C}$ is determined by a classical statistical mechanics model defined on a 3D honeycomb grid of a linear dimension $L$ whose vertices are either empty, $n_{\bf r}=0$, or occupied, $n_{\bf r}=1$. The classical statistical model is defined by the Hamiltonian $H_A = -\sum_{ \left \langle {\bf r},{\bf r'} \right \rangle}n_{\bf r}n_{\bf r'}$, where $\left \langle.,.\right \rangle$ denotes neighboring vertices of the lattice. We simulate the equilibrium distribution of the classical model at an artificially defined temperature $T$ with a geometric cluster Monte-Carlo (MC) approach~\cite{heringa98, heringa98_2}, assuming that a fixed fraction $p$ of sites is occupied. For each computed sample from the equilibrium distribution of the statistical model, we find the largest domain of the occupied sites ($n_{\bf r}=1$) and identify it with the lattice $\mathcal{C}$ upon which the tight-binding Hamiltonian \eqref{eq:Ham} is embedded, and $\mathcal{T}$ is taken as the set of nearest neighbor links between the sites of $\mathcal{C}$.

The above construction  provides us with a controllable parameter $T$ which determines the regularity of the structure of the set $\mathcal{C}$. By varying $T$, we can tune the strength of the structural disorder. Indeed, for $T\to 0$, the attractive interaction ensures that the occupied sites form, in equilibrium, a large single cluster with a smooth surface, as in Fig.~\ref{fig:cartoon_intro} (right). 
On the contrary, one expects a completely random distribution of the 
occupied sites for $T\gg 1$, resulting in the formation of fragmented clusters with irregular boundaries, as in Fig.~\ref{fig:cartoon_intro} (left). An order-disorder phase transition of the classical statistical model at a critical temperature $T_c$, dependent on the fraction $p$ of the occupied sites, can be identified~\cite{suppl}.
A previous work~\cite{Tomasi22} focusing on a similar model in 2D has shown that the order-disorder phase transition coincides with a delocalization-localization transition in the eigenstates of Hamiltonian~\eqref{eq:Ham}, with critical exponent identical to the exponent of the underlying classical statistical model. Here, we focus on the scenario $T>T_c$, setting $T=1$ and examining the properties of eigenstates of~\eqref{eq:Ham} a function of $p$ whose increase results in a gradual decrease of the structural disorder strength.

\paragraph{Localization-delocalization transition in the honeycomb model.}
To reveal the spatial localization in the honeycomb model, we examine the level statistics of $H_{sp}$. We employ shift-and-invert exact diagonalization technique~\cite{Hernandez05,petsc} to find the eigenvalues and eigenvectors of the system, $H_{sp}\ket{\phi_n}=E_n\ket{\phi_n}$, for lattices $\mathcal{C}$ comprising of up to $10^6$ lattice sites. For eigenvalues sorted in ascending order, $E_1<E_2<\dots$, the gap ratio $r$~\cite{Oganesyan07, Atas13} is defined as $r=\langle\min(s_n, s_{n+1})/\max(s_n, s_{n+1})\rangle$, where $s_n=E_{n+1}-E_n$. 
When the eigenstates of $H_{sp}$ are localized, the level statistics are Poissonian, with the average gap ratio $\langle r\rangle = \overline{r}_{PS} \approx 0.386$.
In contrast, for delocalized eigenstates, the level statistics adheres to predictions of random matrix theory, with the average gap ratio $\langle r\rangle =\overline{r}_{GOE} \rangle \approx 0.531$~\cite{Atas13}.
We shall restrict our analysis to eigenstates with energy $E\approx 2.0$, which lie sufficiently away from the singularities in the density of states of $H_{sp}$ associated with eigenstates localized on  lattice sites at the irregular edges of the cluster~\cite{Kosior17, Tomasi22}. Results for other values of $E$ are presented in the supplementary material \cite{suppl}.

\begin{figure}
    \centering
    \includegraphics[width=\linewidth]{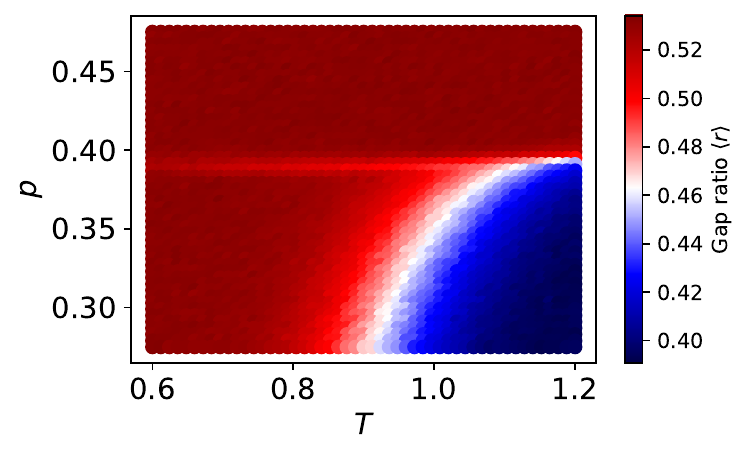}
    \caption{Phase diagram in the $p$ vs $T$ plane showing localized and delocalized phases of the honeycomb model.}
    \label{fig:phase}
\end{figure}

In Fig.~\ref{fig:gap_ratio}a, we plot the average gap ratio $\langle r\rangle$, where the averaging is performed over $100$ eigenstates around $E=2.0$ and over different realization of the lattice $\mathcal{C}$ ($20000$ samples for $L<70$ and $4000$ samples for $L\geq70$), as a function of the occupation fraction $p$ for different system sizes $L$. We observe a transition between localized and delocalized phases in which, with increase of the system size $L$, we find $\langle r \rangle \to r_{PS}$ and $\langle r \rangle \to r_{GOE}$, respectively. The two phases are separated by a critical point at  $p=p_c\approx 0.36$ at which $\langle r \rangle$ remains independent of the system size. 
In the vicinity of the critical point, the gap ratio follows the universal single-parameter scaling form, $r\sim f[(p-p_c)L^{1/\nu}]$, where $f$ is a scaling function and $\nu\approx 1.57$. The value of the critical exponent $\nu$ is consistent with the results for the 3D Anderson model with uncorrelated random~\cite{Slevin18} and quasiperiodic~\cite{luo2022} potentials, suggesting that the transition belongs to the standard Anderson universality class for 3D systems~\cite{Evers08}. Furthermore, we also find evidence for the existence of a mobility edge \cite{suppl}.

The onset of the AL transition as a function of $p$ can be understood as follows. While the structure of the lattice $\mathcal{C}$ is irregular for any $T>T_c$ and $p<1$, the destructive interference effects arising from scattering on these irregularities becomes sufficiently strong to give rise to AL only below $p_c$, i.e., when the strength structural disorder exceeds the critical threshold. We note that one can also observe an AL transition as a function of $T$ if $p$ is held constant to a sufficiently small value, which is similar to the transition observed in~\cite{Tomasi22}. For completeness, we plot a phase diagram in the $p$ vs $T$ plane in Fig.~\ref{fig:phase}, illustrating a sharp border between localized and delocalized phases. The phase diagram clearly shows that having a sufficiently small value of $p$ is strictly necessary in order for the lattice irregularities at high $T$ to induce localization.

\begin{figure}
    \centering
    \includegraphics[width=\linewidth]{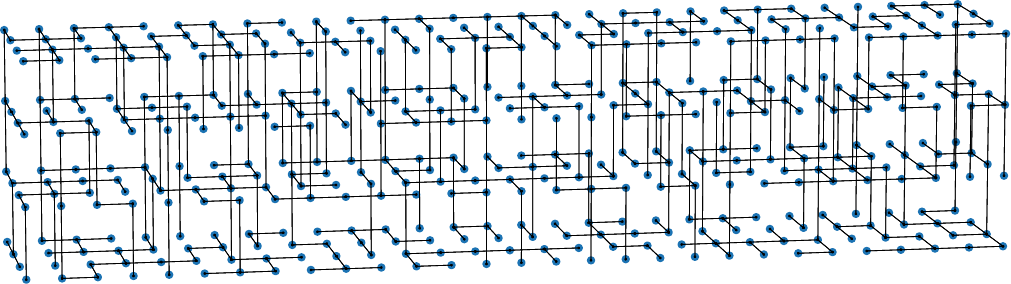}
    \includegraphics[width=\linewidth]{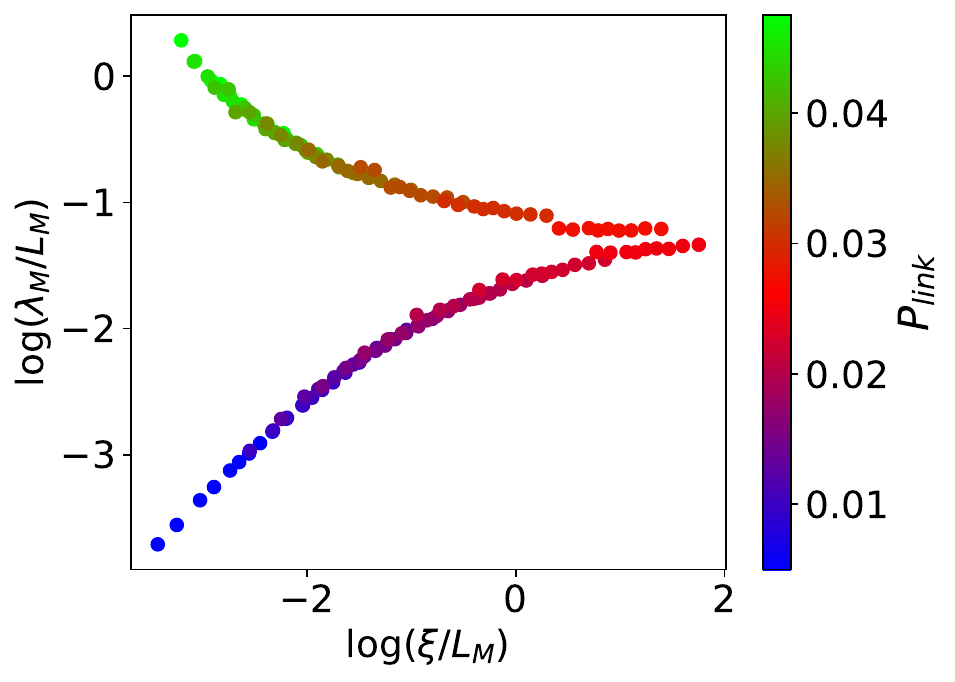}
    \caption{Finite size scaling analysis of the AL transition in the link model. The existence of two branches indicate distinct scaling trend of the localization length $\lambda_M$ below and above $P_{link,c}\approx 0.024$. For $P_{link}>P_{link,c}$, the localization length remains finite and approaches $\xi$ as $L_M\to\infty$ which is characteristic of a localized phase. On the contrary, $\lambda_M$ becomes comparable with $L_M$, thus pointing to a delocalized phase.}
    \label{fig:scaling}
\end{figure}

\paragraph{The link model.}
To verify the robustness of our conclusions about the role of structural disorder for AL transition in 3D lattice models, we now turn to a \textit{link model} with the Hamiltonian~\eqref{eq:Ham} defined on a cubic 3D lattice $\mathcal{C}$ with a set of unidirected links $\mathcal{T}$. The links $\mathcal{T}$ are initially generated as a random spanning tree of the 3D cubic lattice, i.e., a tree graph that includes all vertices of the lattice, with use of Wilson's algorithm~\cite{Wilson96}. 
Subsequently, we add more links to the set $\mathcal{T}$. To that end, for each site $\bf{r}$ of the lattice, we generate all its neighbors $\bf{r}'$ in the 3D cubic lattice geometry, and if the link $(\bf{r},\bf{r'})$ does not belong to $\mathcal T$, it is added to $\mathcal{T}$ with probability $P_{link} \in[0,1]$, giving rise to a lattice schematically represented in Fig.~\ref{fig:scaling}. The probability $P_{link}$ is a control parameter of the link model. For $P_{link}=0$, the lattice is a random spanning tree of the 3D cubic lattice, cutting any of the links results in separating the lattice into two disjoint sublattices. For $P_{link}>0$, the graph contains more links, some of them belonging to loops, while for $P_{link}=1$, the graph becomes the full regular 3D cubic lattice. Analysis of the average gap ratio for the link model reveals properties analogous to the honeycomb lattice, with an AL transition at $P_{link,c} \approx 0.024$, as shown in Fig.~\ref{fig:gap_ratio}b, with the critical exponent  $\nu\approx 1.61$ in agreement with the 3D AL transition class.
Drawing parallel with the honeycomb model, it is interesting to note that increasing $P_{link}$, i.e., adding more links to $\mathcal{T}$, results in the creation of closed loops through which the particles can propagate freely. 
The AL observed in this model thus exemplifies that  transitions induced by structural disorder are generic phenomena that can manifest themselves in many different systems.

In addition to the numerical evidence based on level statistics of $H_{sp}$, we examine the robustness of the AL in the link model employing a renormalized Green's function (RGF) technique introduced in Refs.~\cite{mackinnon85, wurtz88}.
To that end, we consider the link model on a lattice  of dimensions $L_M\times L_M\times L_N$, built by gluing together 2D slices of dimension $L_M\times L_M$ along the horizontal direction, see Fig.~\ref{fig:scaling}~(top). When the transversal dimension, $L_M$, is kept constant, the system is effectively 1D, and, hence, AL for any non-vanishing value of $P_{link}$. We compute the localization length $\lambda_M$ numerically, by solving the RGF's equations, see~\cite{suppl} for details.
The localization length $\lambda_M$ along the horizontal direction is expected to follow a universal scaling of the form $\lambda_M/L_M=F(\xi(P_{link},E)/L_M)$, where $\xi(P_{link},E)$ depends only on $P_{link}$ and the energy $E$ at which the Green's function is calculated. 
Indeed, plotting $\lambda_M/L_M$ vs $\xi/L_M$ (in log-log scale), see Fig.~\ref{fig:scaling}, we observe two distinct curves which can be interpreted as follows. For $P_{link}<P_{link,c}$, the ratio $\lambda_M/L_M$ vanishes with increasing $L_M$ which implies localized eigenstates. On the contrary, for $P_{link}>P_{link,c}$ localization length increases with $L_M$, thus implying the presence of delocalized eigenstates. The RGF analysis qualitatively confirms the conclusion about the AL transition at $P_{link}=P_{link,c}$ in the link model.

\paragraph{Conclusion. }
We have shown that Anderson localization transition can be induced solely by irregularities of 3D lattices, even in the absence of on-site disordered potential. Analyzing statistics of energy levels of tight-binding Hamiltonians defined on irregular 3D lattices, we have shown that the AL transition belongs to the same universality class as the AL induced by the onsite disorder, despite the presence of long-range correlations in the disorder. The same conclusions can be drawn from the study of the participation entropies~\cite{Mace19Multifractal, Liu2024ipr}, which describe the spatial extent of the eigenstates, see~\cite{suppl}. 

The irregularities of 3D lattices play a role of a structural disorder which induces the AL. In the honeycomb model, motivated by complex patterns of graphene platelets, the structural disorder's strength was tuned by changing the fraction $p$ of occupied sites in the underlying classical statistical model. Experimental studies of conductivity of the graphene platelets structures, obtained, e.g., in the process of high-temperature carbonization of charcoal~\cite{Bourke07}, could provide a way of investigating AL. However, we note here that the honeycomb model considered in this work is a much simplified model of carbonized charcoal that incorporates the structural disorder present in such structures, specifically, the disorder arising from the irregular shapes. A possible direction of future research would be to incorporate additional features of these structures, such as the randomness in alignment of the platelets and investigate their consequences. In the link model, the strength of the structural disorder was decreasing when links were added to a random spanning tree of a 3D cubic lattice. Studies of tight-binding models defined on fractal lattices~\cite{Kosior17, RojoFrancas24} provide examples of localization/delocalization phenomena in irregular lattices. A more general understanding of mechanism which determines the strength of the structural disorder is a question open for further investigations.

\begin{acknowledgments}
\paragraph{Acknowledgements.}
S.B. and M.D. acknowledge K.
Jurkiewicz for discussions on the honeycomb lattice
structure. J.Z. acknowledges the discussion with F.
Evers and A. Scardicchio on the nature of Anderson
criticality. ICFO-QOT group acknowledges support
from European Research Council AdG NOQIA;
MCIN/AEI (PGC2018-0910.13039/501100011033,
CEX2019-000910-S/10.13039/501100011033, Plan
National FIDEUA PID2019-106901GB-I00, Plan National
STAMEENA PID2022-139099NB, I00, project funded by
MCIN/AEI/10.13039/501100011033 and by the “European
Union NextGenerationEU/PRTR” (PRTR-C17.I1), FPI);
QUANTERA DYNAMITE PCI2022-132919, QuantERA
II Programme co-funded by European Union’s Horizon
2020 program under Grant Agreement No. 101017733;
Ministry for Digital Transformation and of Civil Service
of the Spanish Government through the QUANTUM
ENIA project call–Quantum Spain project, and by the
European Union through the Recovery, Transformation and
Resilience Plan - NextGenerationEU within the framework
of the Digital Spain 2026 Agenda; Fundació Cellex;
Fundació Mir-Puig; Generalitat de Catalunya (European
Social Fund FEDER and CERCA program; Barcelona
Supercomputing Center MareNostrum (FI-2023-3-0024);
(HORIZON-CL4-2022-QUANTUM-02-SGA PASQuanS2.1,
101113690, EU Horizon 2020 FET-OPEN OPTOlogic, Grant
No. 899794, QU-ATTO, 101168628), EU Horizon Europe
Program NeQST Grant Agreement 101080086–NeQST);
ICFO Internal “QuantumGaudi” project. P.S. acknowledges
fellowship within the “Generación D” initiative, Red.es,
Ministerio para la Transformación Digital y de la Función
Pública, for talent atraction (C005/24-ED CV1), funded by
the European Union NextGenerationEU funds, through
PRTR. The work of J.Z. was funded by the National
Science Centre, Poland, Project No. 2021/03/Y/ST2/00186
within the QuantERA II Programme that has received
funding from the European Union Horizon 2020 research
and innovation programme under Grant Agreement No.
101017733 and the OPUS call within the WEAVE programme
2021/43/I/ST3/01142.
\end{acknowledgments}

\bibliography{ref_21.bib}
\pagebreak
\onecolumngrid
\section{Supplementary material - Anderson localisation induced by structural disorder}

\section{Geometric Monte Carlo Algorithm}
The lattice configurations for the 3D honeycomb model analysed in the main text are generated using a geometric Monte Carlo (GCA) algorithm \cite{heringa98, heringa98_2} as discussed below. We consider an Ising-like nearest-neighbor interaction between the atoms on the lattice of linear dimension $L$:
\begin{equation}
	H = -J\sum_{<{\bf r},{\bf r'}>}n_{\bf r}n_{\bf r'},
\end{equation}
where $n_{\bf r(r')}\in\{0,1\}$ and $\frac{1}{L^3}\sum_{\bf r} n_{\bf r} = p$. The goal of the Monte-Carlo technique is to attain the equilibrium distribution of the above system for an artificially defined temperature $T$ ($J=1, k_B=1$). To this end, we employ the GCA algorithm, which makes use of geometrical symmetries to speed up the Monte-Carlo simulations under the constraint $p={\rm const}$. Consider a symmetry transformation of the lattice $\mathcal{T}\equiv \{{\bf r}\}\to \{{\bf \tilde{r}}\}$, which can be, for example, a reflection along a certain symmetry axis. For a pair of neighboring sites ${\bf r}_a$ and ${\bf r}_b$, we define,

\begin{equation}
	\Delta_{ab}=-\left(n_{{\bf r}_a}n_{{\bf r}_b} + n_{{\bf \tilde{r}}_a}n_{{\bf \tilde{r}}_b} - n_{{\bf r}_a}n_{{\bf \tilde{r}}_b} - n_{{\bf \tilde{r}}_a}n_{{\bf r}_b}\right),
\end{equation}
which calculates the change in \textit{bond energy} associated with swapping the occupation of site ${\bf r}_a$ with its symmetric twin with respect to its neighbor ${\bf r}_b$. The algorithm consists of the following steps:

\begin{enumerate}
	\item Pick a random site ${\bf r}_c$. 
	
	\item Swap $n_{{\bf r}_c}$ and $n_{{\bf \tilde{r}}_c}$.
	
	\item For every nearest-neighbor ${\bf r}_k$ of ${\bf r}_c$, do the following if $\Delta_{ck} > 0$:
	Choose a random number $0\leq z \leq 1$. If $1-\exp[-\Delta_{ck}/T] > z$, swap $n_{{\bf r}_k}$ and $n_{{\bf \tilde{r}}_k}$. Add ${\bf r}_k$ to the stack $S$.
	
	\item Pick a site ${\bf r}_s$ from the stack S. Substitute ${\bf r}_s$ for ${\bf r}_c$ and repeat steps 2 and 3. Remove ${\bf r}_s$ from the stack S.
	
	\item Repeat 4 until S is empty.
\end{enumerate}
The above algorithm guarantees convergence to the equilibrium distribution. As the occupations are always swapped in the above algorithm, the constraint $p={\rm const}$ is always satisfied. In our case, we have randomly chosen reflection along cartesian $x, y$ or $z$ direction as the symmetry transformation after each MC iteration (steps 1-5 is one iteration). The location of the reflection axis is also chosen randomly.

\section{Critical transition with $T$ in the underlying statistical model of honeycomb lattice}

\begin{figure}[h]
	\centering
	\subfigure[]{
		\includegraphics[width=0.4\linewidth]{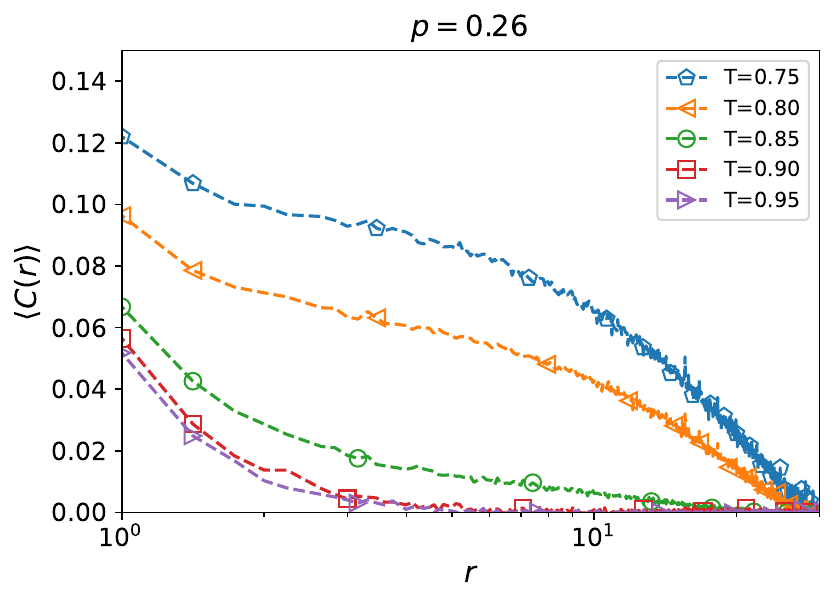}\label{fig:corr_p0p26}}
	\subfigure[]{
		\includegraphics[width=0.4\linewidth]{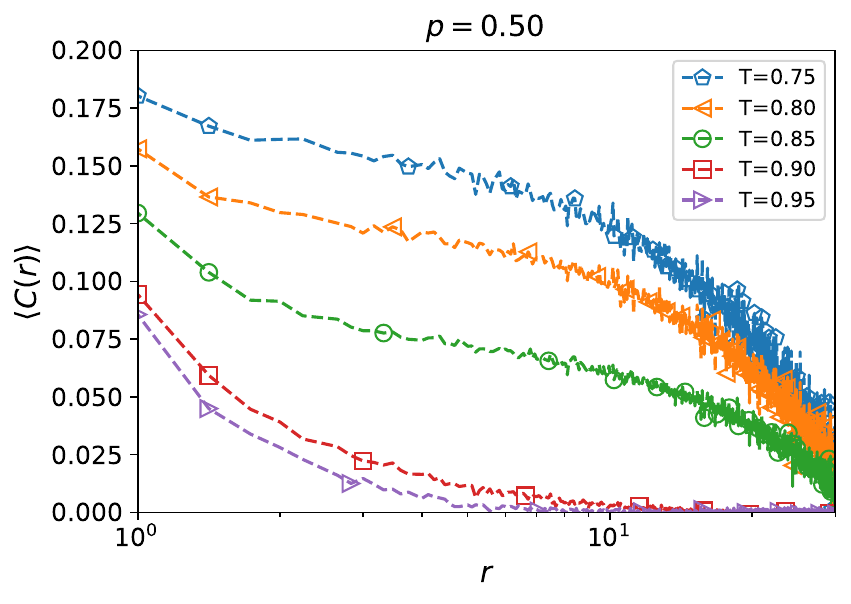}\label{fig:corr_p0p5}}%
	
	\caption{Connected correlation function averaged over disorder realizations at (a) $p=0.26$ and (b) $p=0.50$. In both the cases, the correlation switches from a power-law decay to exponential decay at $T\approx0.85$, demonstrating the presence of a critical structural transition.}
\end{figure}

The 3D honeycomb lattice undergoes a critical transition at temperature $T_c\approx0.85$. To demonstrate it, we calculate the connected correlation function $C(r)=\langle n_{\bf r_a} n_{\bf r_b}\rangle - \langle n_{\bf r_a}\rangle \langle n_{\bf r_b}\rangle$, where $r = |{\bf r_a}-{\bf r_b}|$ is the distance between the sites $a$ and $b$, 
as shown in Figs.~\ref{fig:corr_p0p26} and~\ref{fig:corr_p0p5} for $p=0.26$ and $p=0.5$, respectively. The connected correlation function switches from a power-law decay to exponential decay across $T\approx0.85$, thus showing that  the lattice undergoes a critical structural transition.

\section{Long-range correlated disorder within largest cluster in the Honeycomb Model}

In this section, we show that the Anderson localization transition reported in the honeycomb model occurs in the presence of long-range correlated disorder.  

\begin{figure}[h]
	\centering
	\subfigure[]{
		\includegraphics[width=0.4\linewidth]{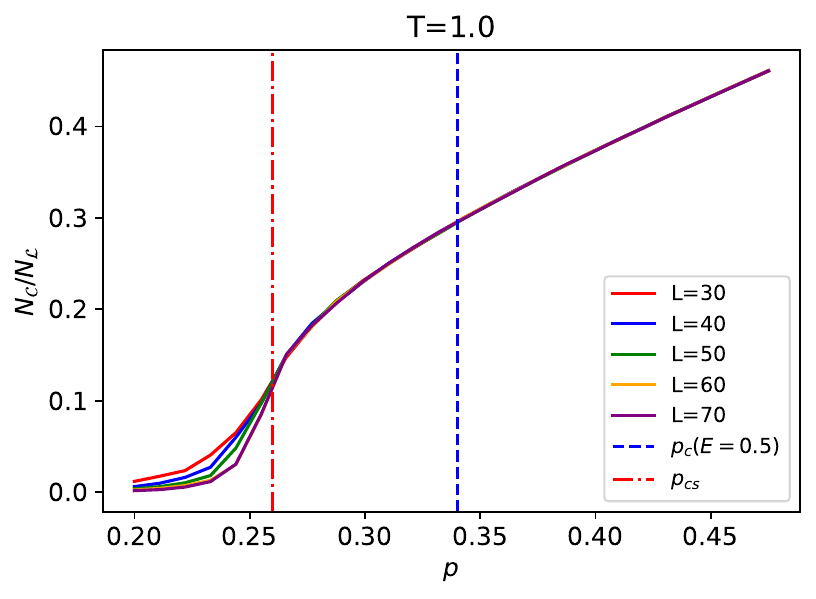}\label{fig:ratio}}
	\subfigure[]{
		\includegraphics[width=0.4\linewidth]{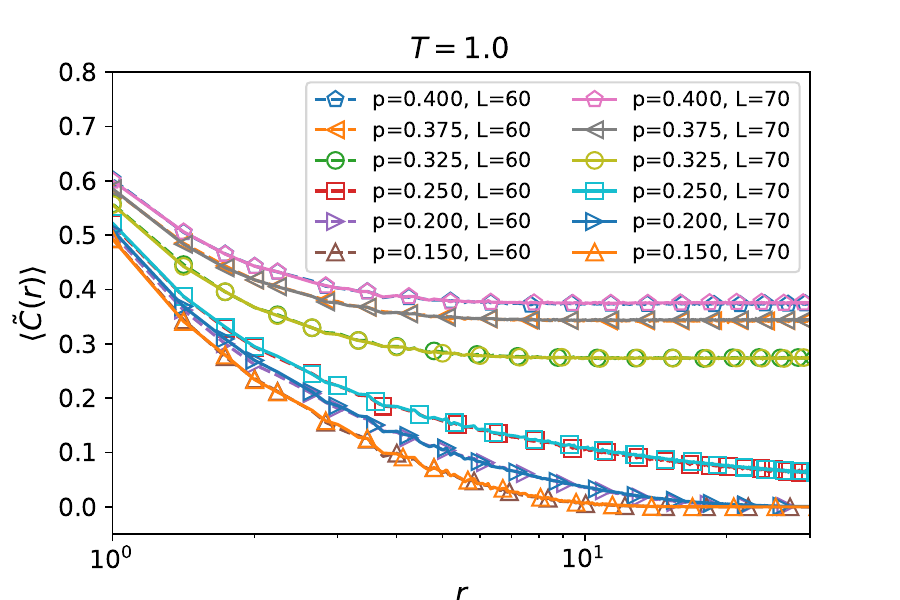}\label{fig:correlation}}%
	\subfigure[]{
		\includegraphics[width=0.4\linewidth]{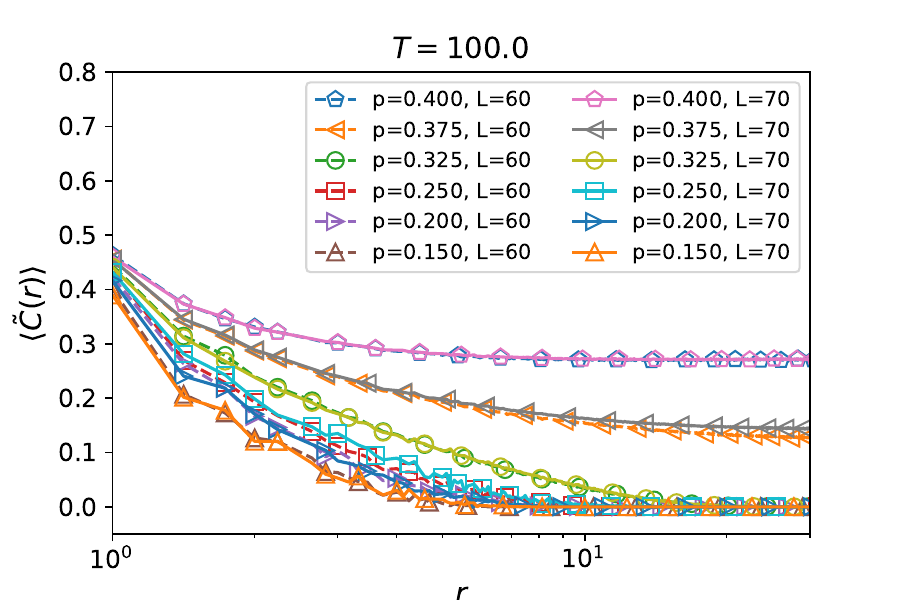}\label{fig:correlation_infty}}
	\caption{(a) Ratio of the largest cluster size $N_\mathcal{C}$ to the lattice size $N_\mathcal{L}$ as as function of $p$ at $T=1.0$. The blue dashed line marks $p_c$ (for $E=0.5$) at which the Anderson localization transition is observed, and the red dash-dotted line marks $p_{cs}$ below which the correlations become short ranged in the underlying statistical model. (b) and (c) show the conditional correlation function, averaged over different disorder realizations, at $T=1$ and $T=100$, respectively. In both cases, the correlation remains finite as long as p is above the corresponding threshold values $p_{cs}$.}
\end{figure}

Firstly, it is important to note that the analysis of the localization properties in the main text is always performed within a single (largest) cluster $\mathcal{C}$ of occupied sites in the lattice $\mathcal{L}$. The largest cluster $\mathcal{C}$ 
size scales proportionally to the lattice size as long as $p>p_{cs}$, where $p_{cs}$ is much smaller than the critical $p_c$ across which we observe the Anderson localization transition. This is shown in Fig.~\ref{fig:ratio} where we plot the ratio of the largest cluster size $N_\mathcal{C}$ to the lattice size $N_\mathcal{L}=L^3$ as a function of $p$ for $T=1.0$. For $p>p_{cs}\approx 0.26$, we see 
that $N_\mathcal{C}/N_\mathcal{L}$
increases monotonically with $p$ and 
the largest cluster contains 
the majority of the occupied sites of the lattice.  We have marked the minimum value of $p_c$ reported in this work (for $E=0.5$) and $p_{cs}$ at $T=1.0$ on Fig.~\ref{fig:ratio} for comparison. 

Secondly, the absence of onsite as well as bond disorder within the cluster implies \textit{perfect long-range correlations} across the whole cluster, which extend throughout the full volume of the lattice as long as  $p>p_{cs}$.
To demonstrate this, we define the variable $\tilde{n}_{\bm r_\alpha}$, such that $\tilde{n}_{\bm r_\alpha}=1$ for $\alpha\in\mathcal{C}$ and $\tilde{n}_{\bm r_\alpha}=0$ for $\alpha\notin\mathcal{C}$. Using this, we define a conditional correlation function $\tilde{C}(r) = \langle \tilde{n}_{{\bm r_\alpha}} \tilde{n}_{{\bm r_\beta}}\rangle$, where $\alpha\in\mathcal{C}\subset\mathcal{L}$, $\beta\in\mathcal{L}$, and $|{\bm r_\alpha}- {\bm r_\beta}| = r$ is the distance between sites $\alpha$ and $\beta$. Fig.~\ref{fig:correlation} shows that the disorder averaged $\tilde{C}(r)$, for $T=1.0$, remains finite at all values of $r$ for $p>p_{cs}$, while it vanishes only for $p<p_{cs}$. The correlation remains finite even at much higher values of $T$, although with an increased $p_{cs}$, as shown in Fig.~\ref{fig:correlation_infty} for $T=100$. This suggests that that correlations persist at $T=\infty$.

\begin{figure}[h]
	\centering
	\subfigure[]{
		\includegraphics[width=0.5\linewidth]{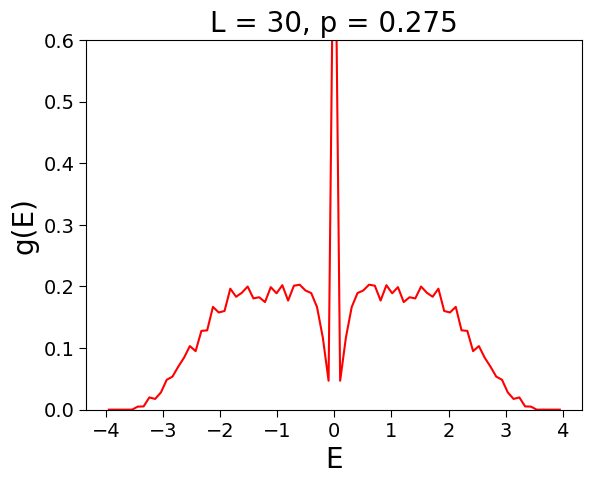}}%
	\subfigure[]{
		\includegraphics[width=0.5\linewidth]{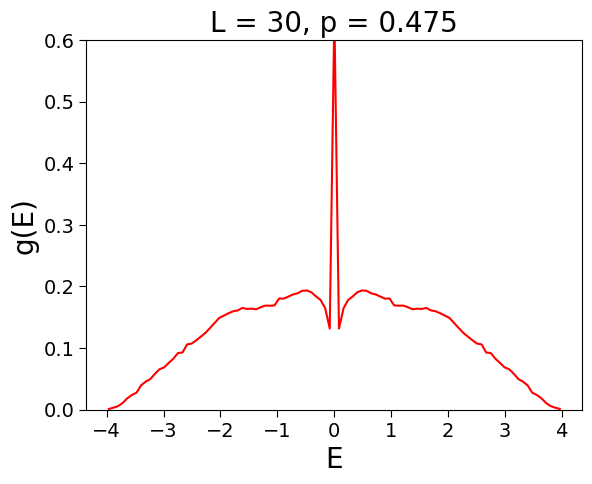}}
	\caption{Density of states $g(E)$ corresponding to the single-particle Hamiltonian $H_{sp}$ of the honeycomb model discussed in the main text for $L=30$, $T=1.0$ and (a) $p=0.275$ and (b) $p=0.475$. The spectrum is symmetric about $E=0$ and has a singularity at $E=0$. The maxima of the energy eigen values vary from $\max |E| \approx 3.5$ at $p=0.275$ to $\max |E| \approx 4.0$ at $p=0.475$.}
	\label{fig:DOE}
\end{figure}

\section{Spectrum of the single-particle Hamiltonian}
\begin{figure}[h]
	\centering
	\subfigure[]{
		\includegraphics[width=0.5\linewidth]{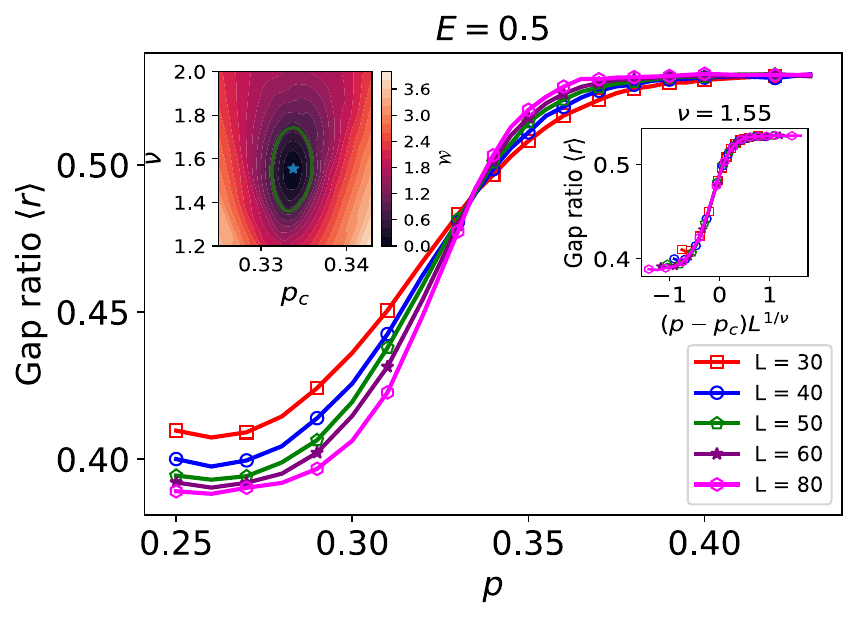}}%
	\subfigure[]{
		\includegraphics[width=0.5\linewidth]{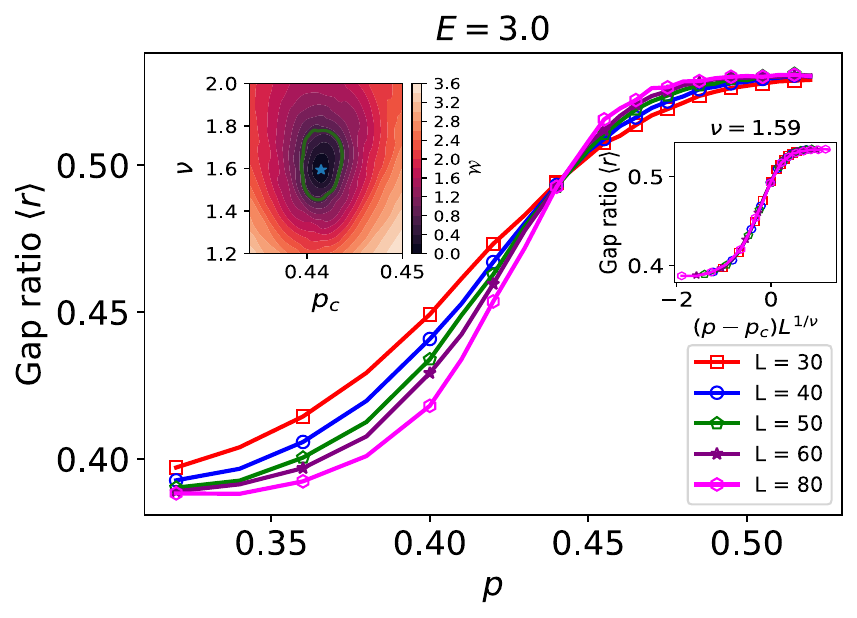}}
	\caption{Gap ratio averaged over 100 nearest energy eigenstates around (a) $E=0.5$ and (b) $E=3.0$, (also averaged over 1000 disorder configurations) as a function of $p$, demonstrating the existence of localisation transition in the honeycomb lattice. The left inset in each of the subfigures shows the cost function (see Eq.~\eqref{eq:supW} of the Supplementary material) used for determining the critical exponent and disorder strength along with the error estimate. The right insets show the scaling collapse of the gap ratio. The critical $p$ increases with $E$; $p_c\approx 0.33$ for $E=0.5$ and $p_c\approx0.44$ for $E=3.0$.}
	\label{fig:trans_diffE}
\end{figure}

In this section, we discuss the nature of the spectrum of the single-particle Hamiltonian $H_{sp}$ of the honeycomb model. In Fig.~\ref{fig:DOE}, we show the density of states (DOS) $g(E)$ across the energy spectrum for the minimum and maximum values of $p$ examined in the main text, i.e., $p=0.275$ and $p=0.475$, respectively. The spectrum is symmetric about $E=0$ where the density of states diverge. In the main text, we choose $E\approx 2.0$ for examining the localisation transition with $p$ to avoid the diverging DOS at $E=0$ and the vanishing DOS at the edges of the spectrum. However, the localisation transition also occurs at other energies, as shown in Fig.~\ref{fig:trans_diffE} for $E=0.5$ and $E=3.0$, respectively.

As discussed in the main text, the gap ratio values are averaged over 100 energy eigenstates nearest to $E=2.0$ for the honeycomb model. This ensures that the eigenstates belong to a small energy window $\Delta E$ with respect to the bandwidth. To illustrate this, we provide data for the relative width $\Delta E/W$ in Table.~\ref{tab:deltaE} for the minimum and maximum values of $p$ examined in the main text and for different system sizes.

\begin{table}[h]
	\centering
	\begin{tabular}{|c|c|c|}
		\hline
		$L$ & $\Delta E/W (p=0.475)$ & $\Delta E/W (p=0.275)$ \\
		\hline
		30 & 0.0152 & 0.0466  \\ 
		40 & 0.0068 & 0.0191 \\
		50 & 0.0035 & 0.0103 \\
		60 & 0.0020 & 0.0060 \\
		70 & 0.0015 & 0.0034 \\
		80 & 0.0010 & 0.0023 \\
		\hline
	\end{tabular}
	\caption{Data showing the relative width of the energy window corresponding to 100 nearest eigenstates around $E=2$ with respect to $W=\max E$.}
	\label{tab:deltaE}
\end{table}

\section{Evidence of mobility edge}
In the main text, we have demonstrated an AL transition in the honeycomb model at $E\approx 2.0$. However, the AL transition also occurs at other energy values. This is demonstrated in Fig.~\ref{fig:trans_diffE} for $E=0.5$ and $E=3.0$. We note that the critical exponent $\nu$ remains invariant (within the error margin) but $p_c$ increases with increasing $E$. This indicates the existence of a mobility edge for $p\in[0.33,0.44]$ in the considered system.

\begin{figure}
	\centering
	\subfigure[]{
		\includegraphics[width=0.5\linewidth]{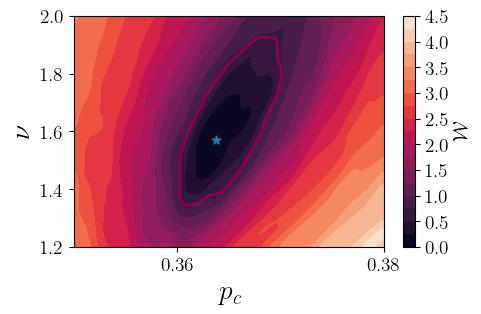}}%
	\subfigure[]{
		\includegraphics[width=0.5\linewidth]{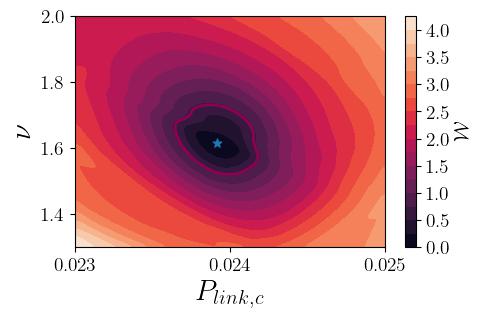}}
	\caption{Cost function defined in Eq.~\ref{eq:supW} for estimating the critical exponent and the critical disorder strength reported in the main text in the case of the (a) honeycomb model and (b) link model.}
	\label{fig:costfunction}
\end{figure}

\section{Estimation of the critical exponents}
In this section, we outline the fitting procedure used for estimating the critical exponents and critical disorder strengths for the AL transition.

The fitting procedure involves minimizing the following cost function:
\begin{subequations}\label{eq:supW}
	\begin{equation}
		\mathcal{W}=\frac{1}{n-2}\sum_{i=2}^{n-1}\mathcal{M}(x_i, y_i, \delta_i |x_{i-1}, y_{i-1}, \delta_{i-1},x_{i+1}, y_{i+1}, \delta_{i+1}), 
	\end{equation}
	with the local weight $\mathcal{M}(x,y,\delta | x', y', \delta', x'', y'', \delta'')$ given by,
	\begin{equation}
		\mathcal{M}=\left(\frac{y-\bar{y}}{\Delta(y-\bar{y})}\right)^2 \quad \bar{y}=\frac{(x''-x)y' - (x'-x)y''}{x''-x'}, \quad \left|\Delta(y-\bar{y})\right|^2 = \delta^2 + \left(\frac{x''-x}{x''-x'}\delta'\right)^2 + \left(\frac{x'-x}{x''-x'}\delta''\right)^2.
	\end{equation}
\end{subequations}
In the above equation, we have the scaling observable $x$ corresponds to $x=(p-p_c)L^{1/\nu}$ for the honeycomb model and $x=(p-P_{link,c})L^{1/\nu}$ for the link model, $y$ corresponds to the averaged gap ratio and $\delta$ is the variance of the observed data. The indices $i$ denote the data points which are arranged such that $x_1<x_2<x_3<\dots$. Minimizing the cost functions amounts to minimizing the distance between 
data curves for different $p$ and $L$ with the ideal minimum value as unity.

In Fig.~\ref{fig:costfunction}, we plot the cost function as a function of the critical disorder strength and exponent $\nu$ for both the honeycomb model and link model. The critical exponent and the critical disorder strength $\alpha =\{p_c, P_{link,c}\}$ are extracted from the minima of the respective plots. The error is identified as the region in the plots for which $\mathcal{W}(\alpha,\nu)<1.3\mathcal{W}(\alpha^*, \nu^*)$.

\begin{figure}
	\includegraphics[width=0.6\columnwidth]{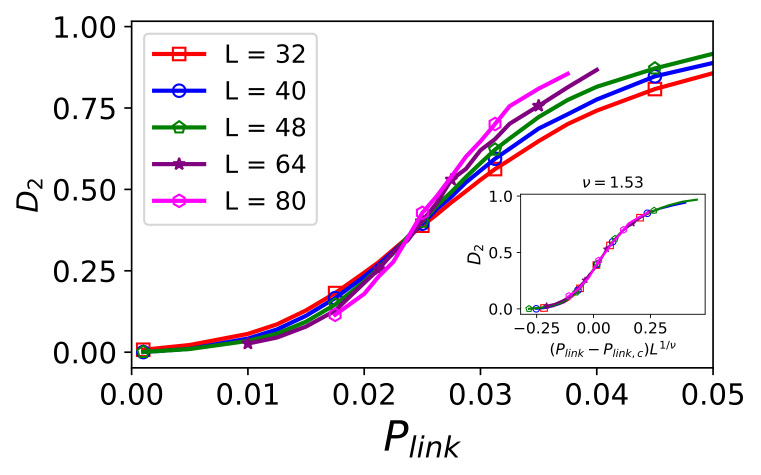}
	\caption{The fractal dimension $D_2$ calculated using Eq.~\ref{eq:sup_Dq} for the link model. An AL transition is identified at $P_{link,c}\approx 0.024$ with $\nu\approx 1.53$, thus corroborating the result obtained from analysing the gap ratio in the main text. The $D_2$ values have been averaged over the same disorder configurations and energy eigenstates that are used for the gap ratio. The $\Delta L$ values chosen for calculating $D_2$ are as follows: $\Delta L = 8$ for $L=32, 40$,  $\Delta L =16$ for $L=48, 64$ and $\Delta L=20$ for $L=80$.}
	\label{fig:supp_dq}
\end{figure}

\section{Analysis of fractal dimension}
The Anderson localization transition is associated with a singular change of the properties of eigenstates, which are delocalized on one side of the transition and become exponentially localized on the other side of (at large disorder). Inverse participation ratios (IPRs), defined as 
\begin{equation}
	I_q =\sum_{\bf r}\left|\braket{\bf r|\phi_n}\right|^{2q},
\end{equation}
enable to assess whether the eigenstate $|\phi_n \rangle$ is localized or delocalized, and, hence, played a vital role in the studies on Anderson transitions~\cite{Thouless74, Kramer93, Evers08}. In the following, we consider participation entropies $S_q$ defined as
\begin{equation}
	S_q =\frac{1}{1-q} \ln\left( I_q \right),
	\label{eq:parti}
\end{equation}
The system size dependence of the participation entropy can be parameterized as 
\begin{equation}
	S_q(L) = D_q \ln( L^d ) + c_q,
	\label{eq:PEL}
\end{equation}
where $D_q$ is fractal dimension~\cite{Halsey86FractalMeasures},  $c_q$ is a sub-leading term, and $d$ is the dimensionality of the system. If the eigenstate $|\phi_n \rangle$ is localized on several sites $| \bf{r} \rangle$, the $S_q$ is independent of $L$ and consequently, the fractal dimension vanishes, $D_q=0$. In contrast, when $|\phi_n \rangle$ is delocalized throughout the whole system, the fractal dimension  is equal to unity, $D_q=1$, and the sub-leading is negative~\cite{Sierant23varying}. Multifractality~\cite{Stanley88} is the intermediate situation when $0<D_q<1$ depends non-trivially on the R\'enyi index $q$, which occurs exactly at the Anderson transition. To extract the fractal dimension, we consider the participation entropy at two system sizes $L$ and $L+\Delta L$, fix the dimensionality of the system as $d=3$ and use Eq.~\eqref{eq:PEL} to extract $D_q$ as
\begin{subequations}\label{eq:sup_Dq}
	\begin{equation}
		D_q = \frac{S_q(L+\Delta L)-S_q(L)}{3\log\left[\frac{L}{L+\Delta L}\right]}.
	\end{equation}
	The participation entropies $S_q(L+\Delta L)$ and $S_q(L)$ are numerically calculated from their definition~\eqref{eq:parti} and averaged over $100$ energy eigenstates $\ket{\phi_n}$ in the vicinity of the target energy and no less than $1000$ disorder realizations. In Fig.~\ref{fig:supp_dq}, we plot $D_2$ as a function of $P_{link}$ for different system sizes $L$. An AL transition can be clearly identified with $P_{link,c}\approx0.024$ and critical exponent $\nu\approx 1.53$, corroborating the results obtained from analysis of the gap ratio in the main text.  
\end{subequations}

\begin{figure}
	\centering
	\includegraphics[width=0.8\linewidth]{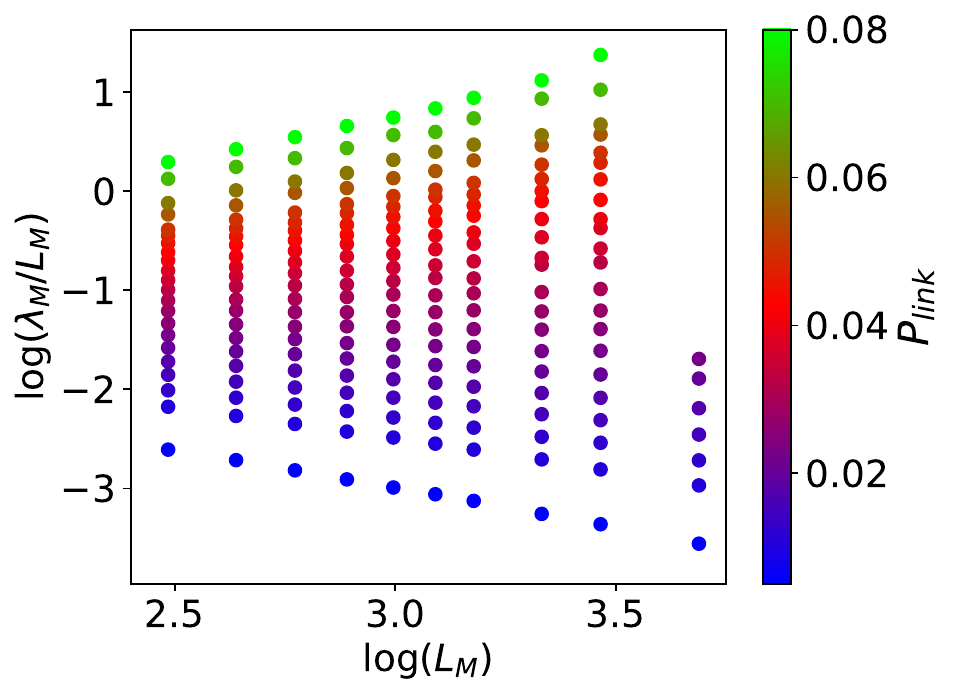}
	\caption{Phase diagram in the $p$ vs $T$ plane showing localized and delocalized phases of the honeycomb model.}
	\label{fig:phase}
\end{figure}

\section{Recursive Green's function method}
We discuss here the recursive Green's function (RGF) technique \cite{MacKinnon81, MacKinnon1983, delande13link} used for calculating the localisation length in the link model. As mentioned in the main text, we visualise that the link model of dimension $L_M\times L_M\times L_N$ is built by glueing together $L_N$ 2D slices of dimension $L_M\times L_M$ along the horizontal direction. Let $H_N$ denote the single-particle Hamiltonian for all $N$ slices. It then follows that,
\begin{equation}
	H_N = H_{N-1} + H_N^{\rm slice} + H_{N-1,N}^{\rm hop} + H_{N,N-1}^{\rm hop},
\end{equation}
where $H_{N-1}$ denotes the Hamiltonian of first $N-1$ slices, $H_N^{\rm slice}$ is the Hamiltonian of the $N^{th}$ slice in the absence of any connection with the preceding $N-1$ slices while $H_{N,N-1}^{\rm hop}$ and $H_{N-1,N}^{\rm hop}$ encapsulate the connections between the $N^{th}$ and $(N-1)^{th}$ slices. 

Let $G^{(N)}_{l,n}$ represent the submatrix connecting layers $l$ and $n$ of the full retarded Green's function of $G^{(N)}=(E+i0^+-H_N)^{-1}$. From Dyson's equations, it follows,
\begin{subequations}
	\begin{equation}
		G^{(N)}_{N,N} = \left(E+i0^+- H_N^{\rm slice} - H_{N,N-1}^{\rm hop}  G^{(N-1)}_{N-1,N-1}H_{N-1,N}^{\rm hop}\right)^{-1},
	\end{equation}
	\begin{equation}
		G^{(N)}_{1,N} = G^{(N-1)}_{1,N-1}H_{N-1,N}^{\rm hop} G^{(N)}_{N,N}.
	\end{equation}
\end{subequations}

Solving the above equations iteratively (see [~\href{https://chaos.if.uj.edu.pl/~delande/Lectures/?transfer-matrix-method,55}{these lecture notes}] for an example of a numerical code) the localisation length along the direction of glueing is then obtained as,
\begin{equation}
	\frac{2}{\lambda_M}=\lim_{N\to\infty}\frac{1}{L_N}\overline{\log{\rm Tr}\left(J^2G^{N}_{1,N}\left[G^{N}_{1,N}\right]^\dagger\right)}
\end{equation}

Having determined the localization length $\lambda_M$ along the horizontal direction, we plot (logarithms of) $\lambda_M/L_M$ as a function of $L_m$ for different values of $\plink$. The correlation length $\xi$ is then subsequently determined by shifting the origin of $\log(L_M)$ to $\log(\xi)$ such that the different curves collapse to form two branches as shown in Fig.~[4] of the main text.

\end{document}